\documentclass[aps,prl,showpacs,twocolumn,superscriptaddress]{revtex4}

\usepackage{graphicx}
\usepackage[ansinew]{inputenc}

\begin{document}

  \bibliographystyle{prsty}

  \title{Conductance and Kondo effect of a controlled single atom contact}
  \author{N. Néel}
  \affiliation{Institut für Experimentelle und Angewandte Physik, Christian-Albrechts-Universität zu Kiel, D-24098 Kiel, Germany}
  \author{J. Kröger} \email{kroeger@physik.uni-kiel.de}
  \affiliation{Institut für Experimentelle und Angewandte Physik, Christian-Albrechts-Universität zu Kiel, D-24098 Kiel, Germany}
  \author{L. Limot}
  \altaffiliation[Present address: ]{Institut de Physique et Chimie des Mat\'{e}riaux de Strasbourg, UMR 7504, Universit\'{e} Louis Pasteur,
  23 rue du Loess, F-67034 Strasbourg, France}
  \affiliation{Institut für Experimentelle und Angewandte Physik, Christian-Albrechts-Universität zu Kiel, D-24098 Kiel, Germany}
  \author{K. Palotas}
  \affiliation{Surface Science Research Centre, University of Liverpool, Liverpool L69 3BX, United Kingdom}
  \author{W.\ A.\ Hofer}
  \affiliation{Surface Science Research Centre, University of Liverpool, Liverpool L69 3BX, United Kingdom}
  \author{R. Berndt}
  \affiliation{Institut für Experimentelle und Angewandte Physik, Christian-Albrechts-Universität zu Kiel, D-24098 Kiel, Germany}
  \date{\today}

  \begin{abstract}
    The tip of a low-temperature scanning tunneling microscope is brought into
    contact with individual Kondo impurities (cobalt atoms) adsorbed on a
    Cu(100) surface. A smooth transition from the tunneling regime to a point
    contact with a conductance of $G\approx\text{G}_0$ occurs. Spectroscopy
    in the contact regime, {\it i.\,e.}, at currents in a $\mu\text{A}$ range
    was achieved. A modified line shape is observed indicating a significant
    change of the Kondo temperature $T_{\text{K}}$ at contact. Model
    calculations indicate that the proximity of the tip shifts the cobalt
    $d$-band and thus affects $T_{\text{K}}$.
  \end{abstract}


  \maketitle

  The concepts of electronic transport in nanometer-sized structures differ
  drastically from those describing macroscopic conductors \cite{nag_03}.
  Measurements of the conductance of metallic nanowires have
  been performed using break junctions, where a thin metal wire is
  mechanically ruptured, or with the scanning tunneling microscope.  The
  results show that chemical properties become important in the extreme case
  of conduction through single atoms or molecules.  In a single-atom contact
  the valence orbitals of the bridging atom act as conductance channels
  \cite{esc_98}, each contributing a fraction of a conductance quantum
  $\text{G}_0=2\,\text{e}^2/\text{h}$ ($\text{e}$: electron charge,
  $\text{h}$: Planck's constant) to the total conductance $G$ according to
  the Landauer formula $G=\text{G}_0\,\sum_i{T_i}$ \cite{esc_97,jcu_98}. The
  transmission $T_i$ of the $i$-th channel is determined by the chemical
  environment. When $T_i$ is spin dependent, the sum extends over both spin
  channels with a maximum conductance of $\text{G}_0/2$ per channel. Complete
  spin polarization and perfect transmission of the relevant channels may
  lead to conductance quantization to integer multiples of $\text{G}_0/2$.
  While this case is not expected from calculations
  \cite{ade_03,aba_04,dja_05,mha_06} there are contradictory experimental
  reports \cite{ton_99,nga_99,mvi_02,cun_04}.

  In break junction experiments usually pure elemental metals are used to
  prepare the contact and the central atom is  most likely the same element
  as the leads. The geometry of the junction and the
  chemical element of the bridging atom are not directly controlled.
  Scanning tunneling microscopy
  (STM) adds more control to the experiment since the bridging
  atom can be a different element than the sample material or
  even a molecule with known orientation \cite{nne_06}.  This opportunity
  is particularly  appealing for magnetic atoms which exhibit the Kondo effect on nonmagnetic substrates.
  By varying the distance of the -- either magnetic or nonmagnetic -- tunneling
  tip from the adsorbed atom (adatom) the many-body Kondo ground state may
  be controllably disturbed.

  Here we report on low-temperature STM experiments on individual Co
  adatoms on Cu(100) surfaces which in spectra of the differential
  conductance ($\text{d}I/\text{d}V$) exhibit a Fano line shape owing to the
  Kondo effect.  When contacting a single atom with the tunneling tip a smooth
  and reproducible transition from the tunneling regime to contact occurs
  with a conductance of $G=I/V\approx\text{G}_0$ ($I$: current, $V$: sample voltage).
  The Cu(100) surface was chosen to provide stable adsorption sites
  for the Co adatoms. Furthermore, spectroscopy in the contact regime, {\it i.\,e.}, at
  currents in a $\mu\,\text{A}$ range is performed without structural
  changes. The line shape is significantly changed at contact and is analysed
  in terms of a modified Kondo temperature $T_{\text{K}}$. On the basis of
  model calculations we propose that, at contact, the Co
  $d$-band shifts and thus affects $T_{\text{K}}$.

  The experiments were performed with a home-made scanning tunneling microscope
  operated at $8\,\text{K}$ and at a base pressure of $10^{-9}\,\text{Pa}$.
  The sample surface was cleaned by argon ion bombardment and annealing.
  Cobalt atoms were deposited onto the Cu(100) surface at $8\,\text{K}$ using
  an electron beam evaporator and an evaporant of $99.99\,\%$
  purity. The tip was prepared {\it in situ} by controlled indentation into
  the substrate until the spectroscopic signature of the Kondo resonance
  appeared as a sharp and reproducible feature in $\text{d}I/\text{d}V$
  spectra. Current-versus-displacement curves were acquired by approaching
  the tip toward the adatom at $45\,\text{\AA}\,\text{s}^{-1}$ and
  simultaneously recording the current.
  Contact formation between the tip and the
  adatom was controllably performed and led to a reproducible contact conductance.
  Subsequent spectroscopy in contact was therefore performed by
  opening the feedback loop at the contact conductances of interest.

  \begin{figure}
    \includegraphics[bbllx=17,bblly=-132,bburx=538,bbury=796,width=75mm,clip=]{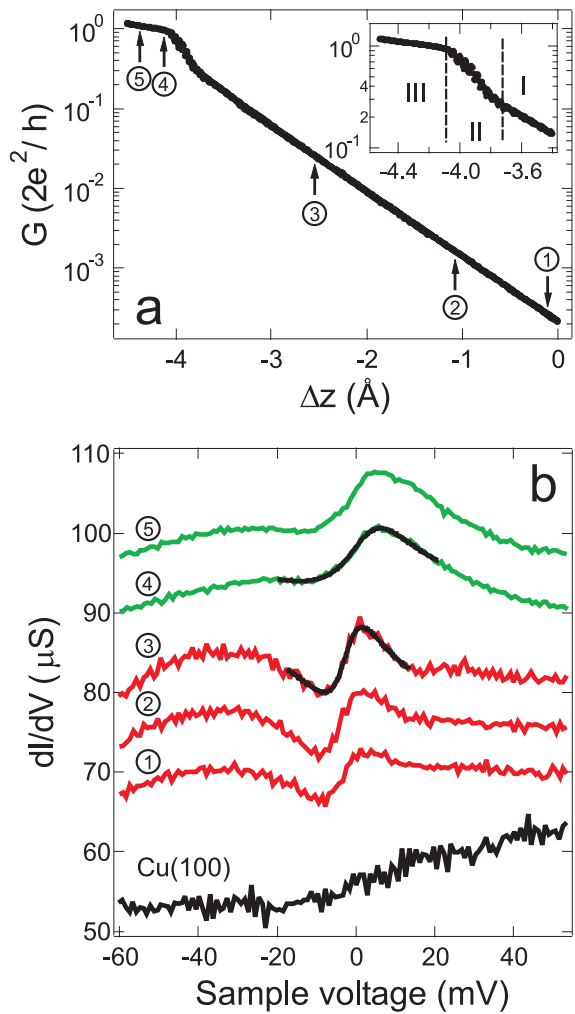}
    \caption[iz_kondo]{(Color online) (a) Conductance $G$ versus tip displacement
    $\Delta z$. Inset: Transition ({\sf II}) from tunneling ({\sf I}) to contact
    ({\sf III}) regime. (b) Spectra of $\text{d}I/\text{d}V$ acquired at
    positions indicated in (a). Lowest curve: Spectrum of Cu(100) at
    $5\,\mu\text{A}$. Curves 1,2,3: Spectra of single Co atom in the
    tunneling regime at $1\,\text{nA}$, $10\,\text{nA}$, $100\,\text{nA}$.
    Curves 4,5: Spectra of single Co atom in the contact regime at
    $5.5\,\mu\text{A}$, $6\,\mu\text{A}$. Solid lines show calculated Fano
    profiles with $q = 1.2$, $T_{\text{K}}= 78\,\text{K}$ (spectrum 3)
    and $q = 2.1 $, $137\,\text{K}$ (spectrum 4).}
    \label{iz_kondo}
  \end{figure}
  Figure \ref{iz_kondo}a shows a typical conductance curve as acquired on
  an individual Co atom on Cu(100). As detailed in Ref.\,\onlinecite{lli_05} the
  linear part of the conductance curve reflects the tunneling regime where
  $G\propto\exp(-1.025\sqrt{\Phi}\Delta z)$ with $\Phi$ the apparent barrier
  height and $\Delta z$ the tip displacement. For this regime denoted by
  {\sf I} in the inset of Fig.\,\ref{iz_kondo}a we find $\Phi\approx 3.5\,\text{eV}$.
  At $\Delta z\approx -3.7\,\text{\AA}$ the slope increases and a continuous
  transition \cite{foot01} (region {\sf II}) occurs from tunneling to contact
  regime (region {\sf III}). The latter is reached at
  $\Delta z\approx -4.1\,\text{\AA}$ and is characterized by a small slope
  (which would correspond to $\Phi\approx 0.3\,\text{eV}$).
  In agreement with the conclusions of Ref.\,\onlinecite{cun_04} for break
  junctions of magnetic metals we find that a single atom Co contact in a
  scanning tunneling microscope does not exhibit a conductance of
  $\text{G}_0/2$.

  In Fig.\,\ref{iz_kondo}b we present spectra of $\text{d}I/\text{d}V$ in the
  tunneling (spectra 1,2,3) and contact (spectra 4,5) regimes.
  The lowest spectrum shows the $\text{d}I/\text{d}V$ signal recorded
  with the same tip in close proximity to clean Cu(100).
  We find the spectroscopic signature of the Kondo resonance around zero sample voltage
  \cite{jli_98,vma_98,ouj_00}
  of Co on Cu(100) \cite{nkn_02}. Intriguingly, this resonance is likewise
  observed in the contact regime. By imaging the surface area prior to and
  after the contact spectroscopy we verified that the tip and the sample
  surface remained unchanged and that the adsorption site of the Co atom was
  not modified. Comparing with spectra from the tunneling regime the current noise
  is appreciably lower at contact. Consequently, $\text{d}I/\text{d}V$
  spectroscopy with the tip of a scanning tunneling microscope in contact
  with an individual atom is feasible in a controlled and reproducible way.

  Comparison of spectra 1--3 and 4,5 reveals a modified line shape at
  contact. Most notably, the line appears broadened compared to the tunneling
  regime. To analyse broadening  of the resonances it is useful to
  describe the spectroscopic signatures by a Fano line shape
  \begin{equation}
    \frac{\text{d}I}{\text{d}V} \propto \frac{(q + \epsilon)^2}{1 + \epsilon^2},
    \label{kondo:fano}
  \end{equation}
  with $\epsilon=(\text{e}V-\epsilon_{\text{K}})/\text{k}_{\text{B}}T_{\text{K}}$
  ($\epsilon_{\text{K}}$: position of Kondo resonance, $\text{k}_{\text{B}}$:
  Boltzmann's constant) and $q$ the asymmetry parameter of the Fano theory
  \cite{ufa_61}. Using Eq.\,(\ref{kondo:fano})
  it is indeed possible to fit the data in Fig.\,\ref{iz_kondo}b. The additional
  width of the Fano feature at contact is reflected by an increased Kondo
  temperature as expected \cite{foot02}.

  \begin{figure}
    \includegraphics[bbllx=0,bblly=0,bburx=750,bbury=384,width=85mm,clip=]{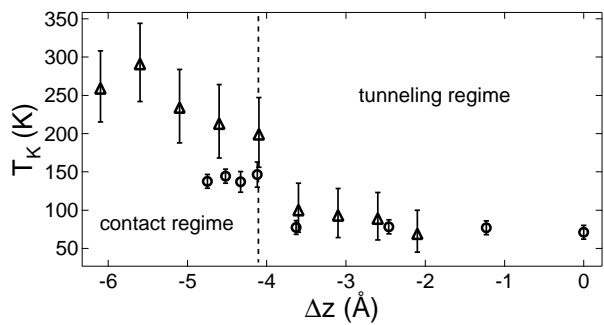}
    \caption[tk]{Kondo temperature $T_{\text{K}}$ versus tip
    displacement. Experimental data are depicted by circles, theoretical data
    are presented by triangles. The dashed line separates tunneling and contact
    regimes.}
    \label{tk}
  \end{figure}
  In Fig.\,\ref{tk} we compare experimentally determined (circles) and calculated
  (triangles) Kondo temperatures. An abrupt change of $T_{\text{K}}$ at a
  displacement of $\Delta z\approx -4.1\,\text{\AA}$ is observed in both data
  sets. For displacements $\Delta z>-4.1\,\text{\AA}$ experimental and
  theoretical Kondo temperatures vary between $70$ -- $100\,\text{K}$. In the
  contact regime, {\it i.\,e.}, $\Delta z<-4.1\,\text{\AA}$, experimental
  Kondo temperatures vary between $140$ and $160\,\text{K}$ while theoretical
  values scatter within $200$ -- $290\,\text{K}$. The abrupt broadening of
  $\text{d}I/\text{d}V$ spectra upon contact formation (see the upper two
  spectra in Fig.\,\ref{iz_kondo}b) can thus be related to a sudden increase
  of the Kondo temperature as depicted in Fig.\,\ref{tk}.

  To determine the origin of the shift in the Kondo temperature we simulated
  the electronic structure of a coupled surface--adatom--tip system with
  standard density functional theory (DFT) \cite{method}. However, a Kondo
  resonance is a genuine many-body effect, the results of groundstate DFT
  simulations can therefore not directly be related to the Kondo temperature.
  Here, we used an approximation going back to the concept of an Anderson
  impurity; in this case the Kondo temperature $T_{\text{K}}$ is described
  by \cite{pwa_61,ash_93,ouj_00}
  \begin{equation}
    \label{kondo}
    T_{\text{K}} \simeq \frac{1}{\text{k}_{\text{B}}}\sqrt{\frac{\Delta U}{\pi}}\exp\left[-\frac{\pi}{\Delta}\left(\left|\frac{1}{\epsilon_d}\right| + \left|\frac{1}{\epsilon_d + U}\right|\right)^{-1}\right],
  \end{equation}
  where $\Delta$ is the crystal field splitting, $\epsilon_d$ the energy
  difference between the occupied $d$-band center and the Fermi level, and
  $U$ the exchange splitting between spin-up and spin-down states. All of
  these variables are readily accessible from groundstate DFT simulations,
  if one assumes that the crystal field splitting $\Delta$, which describes
  the interaction between magnetic states of the impurity and the conducting
  electrons of the metal substrate, is correctly described by DFT simulations
  \cite{ouj_00}, where it shows up as the halfwidth of the spin-up density
  of states (DOS) of the $d$-band. The problem then is transformed
  into finding the center of the occupied (spin-up) $d$-band of the impurity,
  its halfwidth $\Delta$, and the exchange splitting $U$. It also follows
  from this simplified model that the main contributions to a change of the
  Kondo temperature will be a shift of the $d$-band and a broadening or
  narrowing of the band. A similar conclusion was drawn to explain the
  disappearance of the Kondo resonance for Co dimers on Au(111) \cite{che_99}.
  In addition, one expects qualitatively that $\Delta$ will increase with an
  increased coordination number of a magnetic impurity.
  \begin{figure}
    \includegraphics[bbllx=27,bblly=334,bburx=541,bbury=759,width=85mm,clip=]{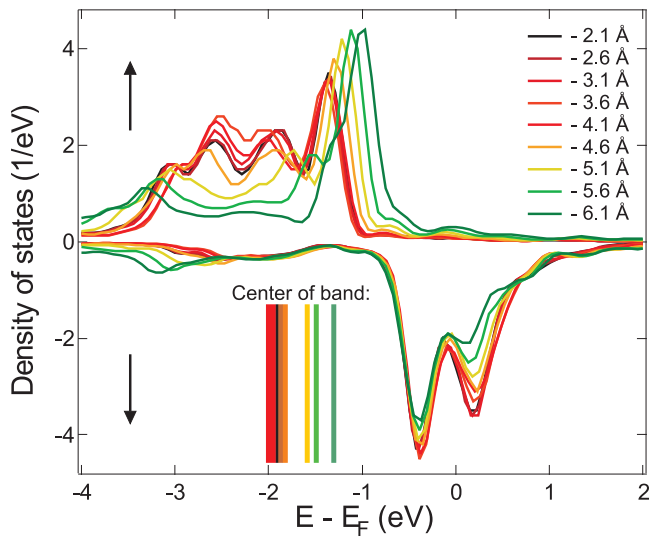}
    \caption[Theory]{(Color online) The density of $d$-states (DOS) with
    varying tip displacement. Spin-up ($\uparrow$) and spin-down ($\downarrow$)
    DOS is plotted as positive and negative values, respectively. The
    $d\uparrow$-band is shifted toward the Fermi energy with decreasing
    displacement. The inset shows the center of the $d$-band calculated
    statistically over the occupied spin-up states.}
    \label{Theory}
  \end{figure}

  In Fig.\,\ref{Theory} we show the spin-polarized density of $d$-states
  at the position of the Co impurity. It can clearly be seen that the main
  changes during an approach occur in the spin-up band of the magnetic impurity.
  Generally speaking, the result of closer proximity of the tip is a shift
  of the band toward the Fermi level and a broadening of the individual peaks.
  Both of these features indicate, qualitatively, that the Kondo temperature
  should increase. We calculated the exchange splitting $U$ by estimating the
  center of the $d$-band for occupied and unoccupied states. The center of the
  spin-up band is shown in the inset of Fig.\,\ref{Theory}. The value we
  obtain for the tunneling regime at a tip displacement of
  $\Delta z = -2.1\,\text{\AA}$ is $U = 2.4\,\text{eV}$, which is slightly
  lower than the value for Co on Au(111) \cite{ivs_98}. This value decreases
  until it reaches $U = 1.9\,\text{eV}$ in the contact regime. In addition,
  the three distinct peaks in the partial DOS merge to a single peak at the
  contact point. This leads to an increase in the crystal field splitting from
  $0.24\,\text{eV}$ in the tunneling regime to $0.40\,\text{eV}$ at contact.
  The crystal field splitting was obtained by fitting the peaks in the DOS
  to a Gaussian. The method is not very precise and the subsequent evaluations
  of the Kondo temperature $T_{\text{K}}$, using Eq.\,(\ref{kondo}), reflect
  the error bar of about $\pm 0.02\,\text{eV}$ in this evaluation. The
  difference between our estimate of $\Delta = 0.24\,\text{eV}$ in the
  tunneling regime and previous work obtaining $\Delta = 0.20\,\text{eV}$
  \cite{ouj_00} may partly be due to this evaluation method. However, also
  in this case the qualitative trend should be that the Kondo temperature
  increases during the tip approach. In Fig.\,\ref{tk} calculated Kondo
  temperatures as a function of the tip displacement are depicted as triangles.
  We observe a close to constant Kondo temperature for $\Delta z > -4.1\,\text{\AA}$
  of about $70$ -- $100\,\text{K}$, in agreement with experimental values.
  The Kondo temperature increases in a step-like fashion if the tip displacement
  decreases below this value. In the contact regime, $\Delta z < -4.1\,\text{\AA}$,
  we find a Kondo temperature of about $200$ -- $290\,\text{K}$, in good
  agreement with the expected value of $300\,\text{K}$ for a magnetic impurity
  in the bulk.
  Due to the increased crystal field splitting with decreasing displacement,
  {\it i.\,e.}, decreasing tip-adatom separation, and decreasing exchange
  splitting between spin-up and spin-down band, the Kondo temperature
  increases during the approach until it reaches the bulk value of about
  $300\,\text{K}$ \cite{ggr_74}.
  However, the simulated values in this case are about $40$ -- $50\,\%$
  above the values obtained in the experiments. While we cannot completely
  rule out that our simple model in this case does not do justice to the
  physical situation, we assume that the deviation is rather due to limitations
  of the simulated system. The surface--adatom--tip system in the simulations
  is very rigid due to the limited number of crystal layers on either side.
  In addition, the tip in the experiments is a metal alloy comprising an
  interface between tungsten and most likely Cu atoms from the surface, while
  it is a single Cu crystal in the simulations. These deviations between the
  experimental and the simulated physical systems may account for the slight
  deviations in the contact regime. However, it has to be stressed that both,
  the absolute values for the Kondo temperature as well as the significant
  change occurring at the displacement of $\Delta z \approx -4.1\,\text{\AA}$
  are very well accounted for in the simulations. Moreover, it is clear that
  the shift of the $d$-band due to the field of the tunneling tip, and the
  reduction of the exchange splitting in this process are the main causes for
  the observed variations.

  In conclusion, $\text{d}I/\text{d}V$ spectroscopy of individual metal
  adatoms in contact with the tip of a scanning tunneling microscope is
  feasible. These experiments will complement measurements using mechanical
  break junctions and offer additional possibilities.  The sample area for
  spectroscopy can be imaged prior to and after contact measurements and
  there is more control over the atom or molecule bridging the electrodes.
  For Co atoms on Cu(100) we observe a modified line shape near the Fermi
  energy which can be described by an increased Kondo temperature $T_{\text{K}}$.
  Model calculations indicate a shift of the Co $d$-band at contact and a
  concomitant change of $T_{\text{K}}$.

  The Kiel authors thank the Deutsche Forschungsgemeinschaft for financial
  support. L. L. thanks F. Gautier for fruitful discussions. K.\ P.\ is
  supported by the EU STREP project RADSAS. W.\ A.\ H.\
  thanks the Royal Society for support by a University Research Fellowship.


\begin{thebibliography}{nnnyys}
\bibitem{foot01} In region {\sf II} the current exhibits significantly
                   increased fluctuations compared to regions {\sf I} and
                   {\sf III}. This may indicate rapid structural changes
                   beyond the time resolution of the experiment.
\bibitem{foot02} We observe a slightly sharper Kondo feature, and, hence,
                   a lower $T_{\text{K}}$ than reported in
                   Ref.\,\onlinecite{nkn_02}.  Generally, the extracted values
                   of $T_{\text{K}}$ depend to some extent on the voltage interval
                   chosen for the fit.
\end{thebibliography}
\end{document}